\documentclass[floafix,prd,showpacs,showkeys,preprintnumbers,nofootinbib,superscriptaddress,11pt]{revtex4-1}
\usepackage[utf8]{inputenc}
\usepackage[sort&compress]{natbib}
\usepackage{ulem}
\usepackage{bm}
\usepackage{times}
\usepackage{amssymb,amsbsy,amsmath,amsfonts}
\usepackage{graphicx}
\usepackage{epstopdf}
\usepackage{float}
\usepackage{color}
\usepackage{morefloats}
\usepackage{rotating}
\usepackage{srcltx}
\usepackage{slashed}
\usepackage{subfigure}
\usepackage{multirow}
\usepackage{verbatim}
\usepackage{hyperref}
\usepackage{lineno}
\usepackage{overpic}
\usepackage{booktabs}
\usepackage{tabularx}

\begin{document}

\title{Studying the heavy quark spin symmetry multiplet of  hadronic molecules  $\bar{D}^{(*)}\Sigma_c^{(*)}$ in  the three-body decays of $\bar{D}^{(*)}\Lambda_c \pi$  }

\author{Ming-Zhu Liu}
\affiliation{
Frontiers Science Center for Rare Isotopes, Lanzhou University,
Lanzhou 730000, China}
\affiliation{ School of Nuclear Science and Technology, Lanzhou University, Lanzhou 730000, China}

\author{Ya-Wen Pan}
\affiliation{School of Physics, Beihang University, Beijing 102206, China}
\affiliation{Research Center for Nuclear Physics (RCNP), Ibaraki, Osaka 567-0047, Japan}

\author{Li-Sheng Geng}
\email[Corresponding author:]{lisheng.geng@buaa.edu.cn}
\affiliation{School of Physics, Beihang University, Beijing 102206, China}
\affiliation{Beijing Key Laboratory of Advanced Nuclear Materials and Physics, Beihang University, Beijing 102206, China}
\affiliation{Peng Huanwu Collaborative Center for Research and Education, Beihang University, Beijing 100191, China}
\affiliation{Southern Center for Nuclear-Science Theory (SCNT), Institute of Modern Physics, Chinese Academy of Sciences, Huizhou 516000, China}

\begin{abstract}

The decay behavior of an exotic state can be used to probe its internal structure. We note that  
the hidden-charm pentaquark states, $P_{\psi}^{N}(4312)$, $P_{\psi}^{N}(4440)$, and $P_{\psi}^{N}(4457)$, have only been
observed in the $J/\psi p$ channel.  In this work,  we employ the effective Lagrangian approach to systematically investigate the two-body and three-body decays of the heavy quark spin symmetry multiplet of hadronic molecules  $\bar{D}^{(*)}\Sigma_c^{(*)}$. Our results show that the partial decay widths of the hidden-charm pentaquark molecules into $\bar{D}^{(*)}\Lambda_c \pi$ are sizable so that  $\bar{D}^{(*)}\Lambda_c \pi$  are promising channels to search for them, which can help clarify their molecular nature.               

\end{abstract}


\maketitle

\section{Introduction}

In the past 20 years, many states beyond mesons made of a pair of quark and anti-quark and baryons made of three quarks in the conventional quark model, also named exotic states,  have been discovered experimentally, while their nature is still controversial~\cite{Chen:2016qju,Hosaka:2016ypm,Lebed:2016hpi,Oset:2016lyh,Esposito:2016noz,Dong:2017gaw,Guo:2017jvc,Olsen:2017bmm,Ali:2017jda,Karliner:2017qhf,Guo:2019twa,Brambilla:2019esw,Liu:2019zoy,Meng:2022ozq,Liu:2024uxn}.  Among them, the pentaquark states are one of the major categories, which provide a nice platform to understand the non-perturbative property of QCD and study the hadron structure.  In 2015, two hidden-charm pentaquark states $P_{\psi}^{N}(4380)$ and  $P_{\psi}^{N}(4450)$  were observed by the LHCb Collaboration in the $J/\psi p$ invariant mass distributions of  the $\Lambda_{b}\to J/\psi p K$ decay~\cite{Aaij:2015tga}.
Four years later, with a larger data sample in the same process, they found that the original  $P_{\psi}^{N}(4450)$   state splits into two states,  $P_{\psi}^{N}(4440)$ and  $P_{\psi}^{N}(4457)$, and a new state $P_{\psi}^{N}(4312)$ emerges~\cite{Aaij:2019vzc}.   The internal structures of these three pentaquark states have been extensively studied theoretically. Although the $\bar{D}^{(*)}\Sigma_c$ molecular interpretation for $P_{\psi}^{N}(4312)$,  $P_{\psi}^{N}(4440)$,  and  $P_{\psi}^{N}(4457)$ is the most popular~~\cite{Liu:2019tjn,Chen:2019asm,He:2019ify,Chen:2019bip,Xiao:2019aya,Sakai:2019qph,Yamaguchi:2019seo,He:2019rva,Liu:2019zvb,Valderrama:2019chc,Meng:2019ilv,Du:2019pij,Ling:2021lmq,Dong:2021juy,Ozdem:2021ugy,Xing:2022ijm,Pan:2022xxz,Zhang:2023czx,Pan:2022whr,Liu:2023wfo,Pan:2023hrk,Liu:2023wfo}, there exist other explanations, e.g.,  hadro-charmonia~\cite{Eides:2019tgv}, compact pentaquark states~\cite{Ali:2019npk,Ali:2019clg,Wang:2019got,Cheng:2019obk,Weng:2019ynv,Zhu:2019iwm,Pimikov:2019dyr,Ruangyoo:2021aoi}, 
virtual states~\cite{Fernandez-Ramirez:2019koa}, triangle singularities~\cite{Nakamura:2021qvy}, and cusp effects~\cite{Burns:2022uiv}.

Since  $P_{\psi}^{N}(4312)$,  $P_{\psi}^{N}(4440)$,  and  $P_{\psi}^{N}(4457)$ lie close to the $\bar{D}^{(*)}\Sigma_c$ mass thresholds,  they can be nicely arranged into a heavy quark spin symmetry (HQSS) multiplet of  $\bar{D}^{(*)}\Sigma_c^{(*)}$ molecules~\cite{Liu:2019tjn,Xiao:2019aya,Du:2019pij,Liu:2019zvb,Yamaguchi:2019seo}. Assuming  $P_{\psi}^{N}(4440)$ and  $P_{\psi}^{N}(4457)$ as the $\bar{D}^*\Sigma_c$ bound states, there are two alternatives for their spin assignments in the effective field theory (EFT)~\cite{Liu:2019tjn}, namely,  in Scenario A, the spin of  $P_{\psi}^{N}(4440)$ and $P_{\psi}^{N}(4457)$ are  $J^P=1/2^-$ and $J^P=3/2^-$, while they are $J^P=3/2^-$ and $J^P=1/2^-$ in Scenario B. The key issue is the uncertainty in the $J^P=1/2^-$ $\bar{D}^*\Sigma_c$  and $J^P=3/2^-$ $\bar{D}^*\Sigma_c$ potentials. According to the time-honored One-Boson-Exchange (OBE) model,  the $J^P=1/2^-$ $\bar{D}^*\Sigma_c$ potential is more attractive than the $J^P=3/2^-$ $\bar{D}^*\Sigma_c$ potential~\cite{Chen:2019asm,He:2019ify}, but their order reverses once the delta potentials of the OBE model are ignored~\cite{Liu:2019zvb}. The local hidden gauge approach~\cite{Xiao:2019aya} and the lattice QCD simulations~\cite{Xing:2022ijm}  found that the $J^P=1/2^-$ $\bar{D}^*\Sigma_c$ and $J^P=3/2^-$ $\bar{D}^*\Sigma_c$ potentials are almost degenerate. One can see that the spins of   $P_{\psi}^{N}(4440)$ and  $P_{\psi}^{N}(4457)$ can not be determined by studying the $J^P=1/2^-$ $\bar{D}^*\Sigma_c$  and $J^P=3/2^-$ $\bar{D}^*\Sigma_c$ potentials. Based on the proposal that the $\bar{D}^{(*)}\Sigma_c^{(*)}$ system is related to the $\Xi_{cc}^{(*)}\Sigma_c^{(*)}$ system by the Heavy Antiquark Diquark Symmetry (HADS),  we predicted the mass spectrum of the $\Xi_{cc}^{(*)}\Sigma_c^{(*)}$ system in two scenarios~\cite{Pan:2019skd}. The sign of the mass splitting between $J^P=0^+$ $\Xi_{cc}\Sigma_c$ and $J^P=1^+$ $\Xi_{cc}\Sigma_c$ is opposite in the two scenarios, which is likely to be determined by the lattice QCD simulation of the $\Xi_{cc}\Sigma_c$ doublet~\cite{Junnarkar:2019equ}. Recently,  it was claimed that Scenario A is more favored in the neural network approach~\cite{Zhang:2023czx}. 

In addition to the masses of the pentaquark states, studying their production and decay processes is key to verifying their molecular nature as well as fixing the spins of  $P_{\psi}^{N}(4440)$ and  $P_{\psi}^{N}(4457)$.
   In Ref.~\cite{Wu:2019rog}, the authors adopted the triangle diagram mechanism to study the hidden-charm pentaquark productions in the $\Lambda_b$ decays,  which are difficult to be verified experimentally due to the unknown branching fractions of the decays $P_c \to J/\psi p$ and heavily suppressed weak decays $\Lambda_b \to D_s^{(*)} \Sigma_c$~\cite{Falk:1992ws,Gutsche:2018utw}. Using the same approach, we selected the Cabibbo-favored weak decays $\Lambda_b \to D_s^{(*)} \Lambda_c$ to produce the pentaquark states in the $\Lambda_b$ decays, but can not obtain a consistent description of their production rates due to the uncertainties in the branching fractions of the decays $P_c \to \bar{D}^{(*)}\Lambda_c$~\cite{Pan:2023hrk}.  Moreover, the hidden-charm pentaquark state productions in other $b$-hadron decays are analyzed by the topological diagram approach~\cite{Xing:2021yid,Han:2023oac}, which provide likely processes to search for the pentaquark states~\cite{Hsiao:2015nna}. Very recently, the LHCb Collaboration measured the branching fractions of the three-body weak decays $\Lambda_b \to \bar{D}^{(*)}\Lambda_c K$ and $\Lambda_b \to \bar{D}^{(*)}\Sigma_c^{(*)} K$~\cite{LHCb:2023eeb,LHCb:2024fel}, which indicate that producing the pentaquark states via the final-state interactions  $\bar{D}^{(*)}\Sigma_c^{(*)}$ of the three-body weak decays  $\Lambda_b \to \bar{D}^{(*)}\Sigma_c^{(*)} K$ is likely.

In Refs.~\cite{Kuang:2020bnk,Burns:2022uiv},   $P_{\psi}^N(4457)$ is interpreted as a cusp effect. To distinguish whether a pentaquark state is a genuine state or kinetical effect,  observing it in the photoproduction process can be a key~\cite{Wang:2015jsa,HillerBlin:2016odx,Meziani:2016lhg,Karliner:2015voa,Wang:2019krd,Wu:2019adv}. The GlueX  Collaboration did not observe significant evidence for the hidden-charm pentaquark states in the $\gamma p \to J/\psi p$ process~\cite{GlueX:2019mkq}.      Moreover, Li et al. suggested that the hidden-charm pentaquark state can be produced in the $e^+e^-$ collisions~\cite{Li:2017ghe},   and Voloshin proposed to search for the hidden-charm pentaquark state in the $J/\psi p$ or $\eta_c p$  mass distribution in antiproton-deuteron collisions~\cite{Voloshin:2019wxx}. Recently, the Belle Collaboration found no evident signal for $P_{\psi}^{N}(4312)$,  $P_{\psi}^{N}(4440)$,  and  $P_{\psi}^{N}(4457)$  in the $J/\psi p$ mass distribution of the processes $e^+e^- \to \Upsilon(1S,2S) \to J/\psi p+ $anything~\cite{Belle:2024mcb}. Based on Monte Carlo simulations, the inclusive production rates of these pentaquark states are estimated for proton-proton collisions~\cite{Chen:2021ifb,Ling:2021sld} and electron-proton collisions~\cite{Shi:2022ipx}.  Very recently, the LHCb Collaboration found no significant signal of $P_{\psi}^{N}(4312)$,  $P_{\psi}^{N}(4440)$,  and  $P_{\psi}^{N}(4457)$ in the open-charm decay channels in prompt productions~\cite{LHCb:2024pnt}. In addition to observing the hidden-charm pentaquark states in the $\Lambda_b$ decays in proton-proton collisions~\cite{Aaij:2019vzc}, no pentaquark state has been observed in other processes.

Studying the decays of the pentaquark states provides another important probe into their internal structure~\cite{Voloshin:2019aut}. The decay modes of the hidden-charm pentaquark states are complex due to many decay channels like  $X(3872)$~\cite{Li:2019kpj}. From the phase space perspective, the two-body and three-body strong decay modes contribute dominantly to the widths of the pentaquark states.   
In Refs.~\cite{Xiao:2019aya,Xiao:2020frg}, the authors adopted the local hidden gauge approach to study the hidden-charm pentaquark molecules decaying into  
$\bar{D}^{(*)}\Lambda_c$ and $J/\psi(\eta_c) p$, finding that the partial decay  widths of $P_{c} \to J/\psi p (\eta_c p)$ are larger than those of $P_{c} \to \bar{D}^{(*)}\Lambda_c$, in contrast with the OBE model~\cite{Lin:2019qiv,Yamaguchi:2019seo, He:2019rva, Burns:2021jlu}. It is well accepted that the OBE model works well with light meson exchanges but is not verified for heavy meson exchanges, especially when both heavy and light meson exchanges are allowed. This indicates our lack of understanding of hadron-hadron interactions. In Refs.~\cite{Lin:2019qiv,Xie:2022hhv}, the authors estimated that the three-body partial decay widths of  $P_{\psi}^{N}(4312)$,  $P_{\psi}^{N}(4440)$,  and  $P_{\psi}^{N}(4457)$ are small, while the widths of their four HQSS partners could be several MeV. In Ref.~\cite{Du:2021fmf}, Du et al., found that the three-body cut $\bar{D}\Lambda_c \pi$ plays important role in describing the widths of $\bar{D}\Sigma_c^*$ molecule.   
Very recently, the LHCb Collaboration searched for $P_{\psi}^{N}(4312)$,  $P_{\psi}^{N}(4440)$,  and  $P_{\psi}^{N}(4457)$ in the open-charm channels $\bar{D}^{(*)}\Lambda_c$ and $\bar{D}^{(*)}\Lambda_c\pi$, while no significant signal is observed~\cite{LHCb:2024pnt}, which indicates that in addition to the decays $P_c \to J/\psi p$, no other decay channel has been observed experimentally.  In this work, we employ the contact-range EFT   to systematically study the two-body and three-body decays of the hidden-charm pentaquark states within the HQSS multiplet hadronic molecular picture, which is expected to yield more information on their decay behaviors.

This work is organized as follows. We first
briefly introduce the tree-level Feynman diagrams for the hidden-charm pentaquark states as the $\bar{D}^{(*)}\Sigma_c^{(*)}$ molecules decaying into $\bar{D}\Lambda_c\pi$ and $\bar{D}^*\Lambda_c\pi$, and the effective Lagrangian approach in Sec.~II. The numerical results and discussions are given in Sec.~III, followed by a summary in the last section.

\section{ THEORETICAL FORMALISM }
\label{sec:Sec2}

\begin{figure}[ttt]
\begin{center}
\begin{overpic}[scale=0.65]{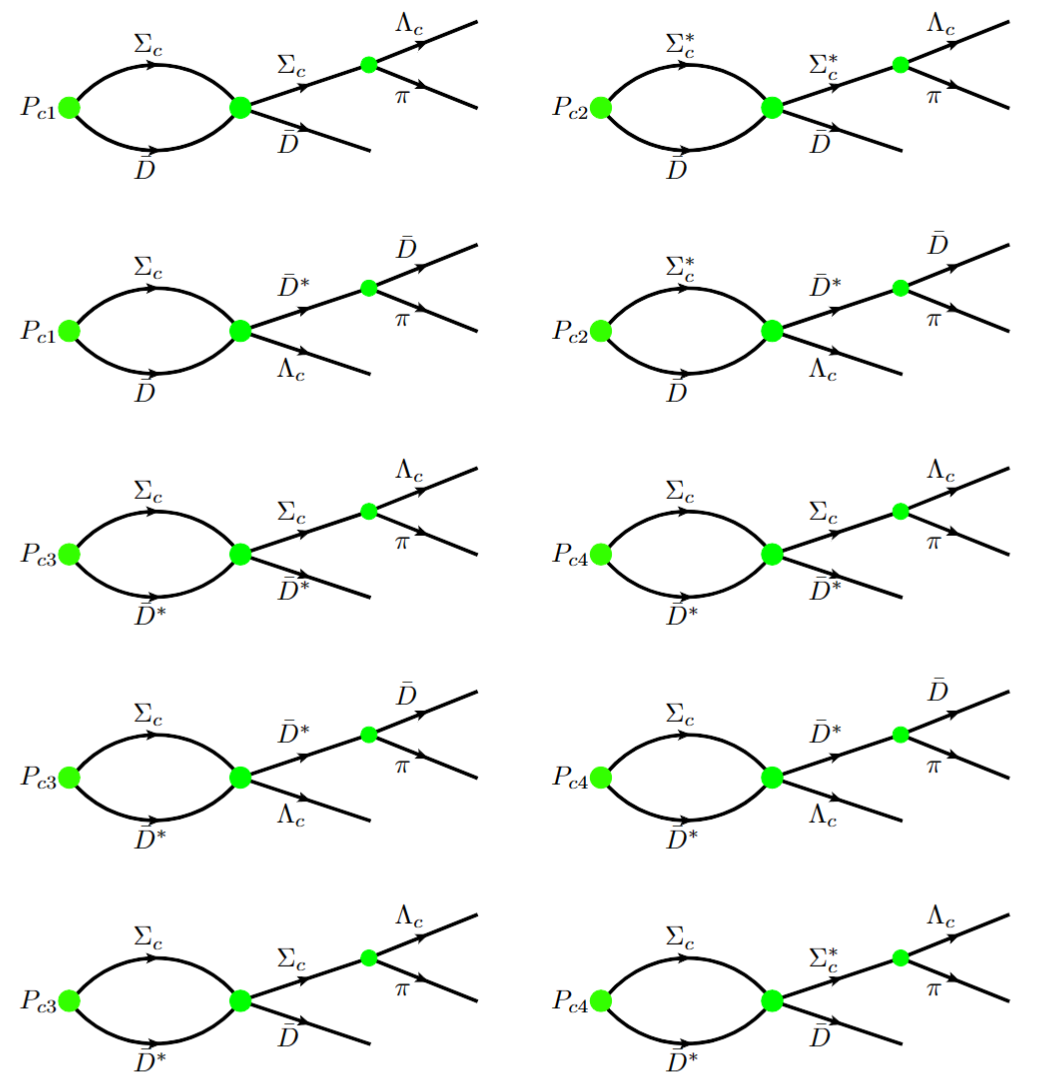}
\end{overpic}
\caption{ Tree-level diagrams of  $P_{c1}$, $P_{c2}$, $P_{c3}$, and $P_{c4}$ predominantly generated by the $\bar{D}^{(*)}\Sigma_c$ potentials decaying  into $\bar{D}\Lambda_c \pi$ and $\bar{D}^*\Lambda_c \pi$.       }
\label{pc14}
\end{center}
\end{figure}

\begin{figure}[ttt]
\begin{center}
\begin{overpic}[scale=0.65]{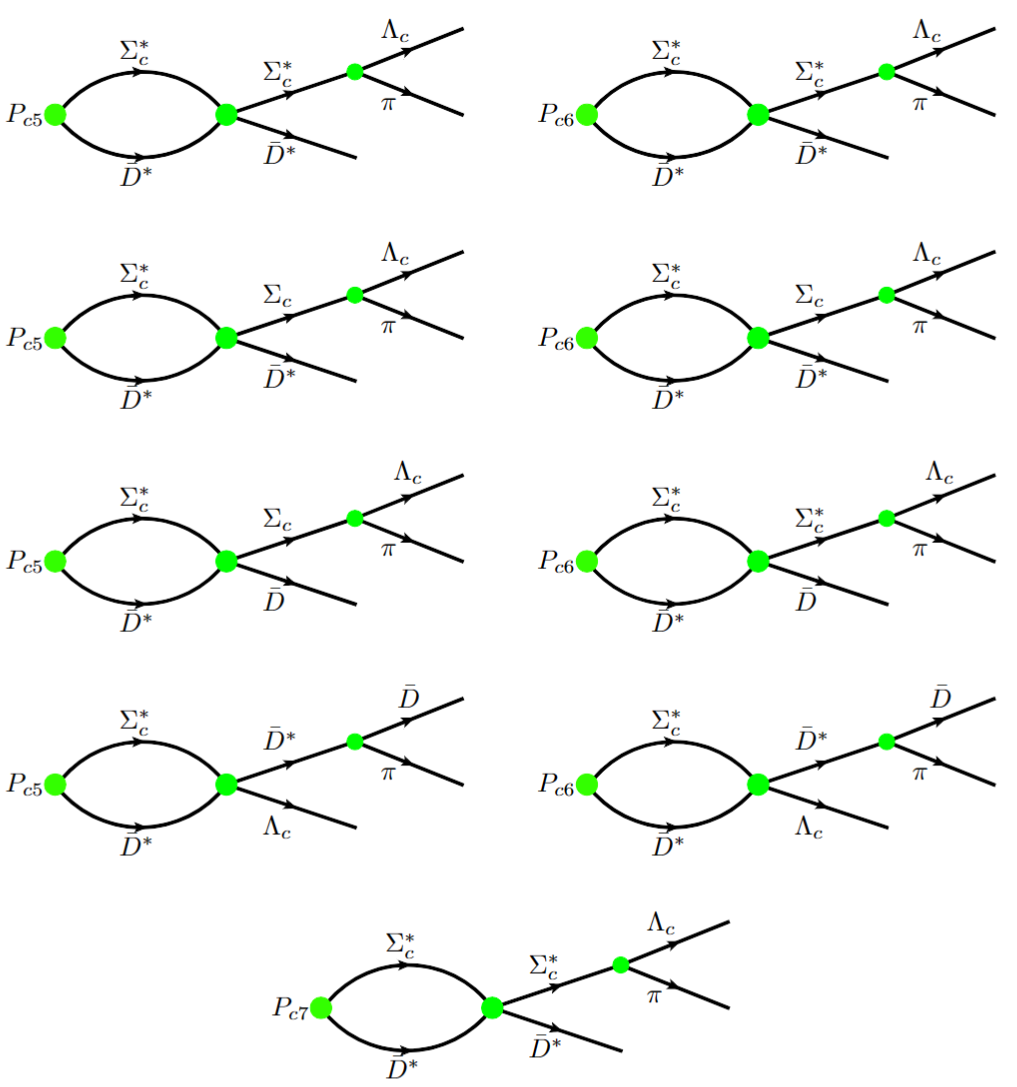}
\end{overpic}
\caption{Tree-level diagrams of   $P_{c5}$, $P_{c6}$, and $P_{c7}$ predominantly  generated by the $\bar{D}^{*}\Sigma_c^*$ potentials decaying  into $\bar{D}\Lambda_c \pi$ and $\bar{D}^*\Lambda_c \pi$.   }
\label{pc57}
\end{center}
\end{figure}

 Assuming  $P_{\psi}^{N}(4312)$, $P_{\psi}^{N}(4440)$, and {$P_{\psi}^{N}(4457)$} as bound states of $\bar{D}^{(*)}\Sigma_c$ 
, four more bound states of $\bar{D}^*\Sigma_c^{(*)}$ are predicted considering HQSS, resulting in a complete HQSS multiplet of hadronic molecules~\cite{Liu:2019tjn}.  However, due to the undetermined spin of $P_{\psi}^{N}(4440)$ and {$P_{\psi}^{N}(4457)$}, we usually employ two scenarios to investigate the hidden-charm pentaquark states.    
In the following, we denote the seven hidden-charm pentaquark hadronic molecules by $P_{c1}$ $\dots$ $P_{c7}$ following the order of Scenario A of Ref.~\cite{Liu:2019tjn}. 
This work mainly focuses on the three-body decay modes of the hidden-charm pentaquark states.
One can classify two kinds of three-body decay modes in the molecular picture.   The first proceeds  via    $\bar{D}^{(*)}\Sigma_{c}^{(*)}$ elastically scattering into  $\bar{D}^{(*)}\Sigma_{c}^{(*)}$ and then  
$\Sigma_c^{(*)}$ baryons decaying into $\Lambda_c \pi$,  and the second  via   $\bar{D}^{(*)}\Sigma_{c}^{(*)}$  inelastically  scattering  into $\bar{D}^{(*)} \Lambda_c(\Sigma_c^{(*)})$ and then  the secondary decays of $\bar{D}^* \to \bar{D}\pi$ and $\Sigma_{c}^{(*)} \to \Lambda_c \pi$ as shown in  Fig.~\ref{pc14} and Fig.~\ref{pc57},  { which are factorized into two processes, e.g., the scattering and secondary decay processes. The loop diagrams in  Fig.~\ref{pc14} and Fig.~\ref{pc57} represent the scattering process, including elastic scattering and inelastic parts, described by a unitary coupled-channel amplitude.  In the present work,  we condense the loop diagram into a vertex,  which is equal to simplifying Fig.~\ref{pc14} and Fig.~\ref{pc57} as tree diagrams. The effective Lagrangians describe such effective vertices.      }   


 In such a picture, the contributions of the two-body decay channels of $\bar{D}^*\Lambda_c$, $\bar{D}^{(*)}\Sigma_c$ and $\bar{D}\Sigma_c^{*}$ to the widths of the pentaquark states are absorbed into the three-body decay channels of $\bar{D}\Lambda_c\pi$ and $\bar{D}^*\Lambda_c\pi$. Therefore,   the hidden-charm pentaquark hadronic molecules have the following decay modes:   two-body decay modes of $\bar{D}\Lambda_c$, $\eta_c p$, and $J/\psi p$  and three-body decay modes of $\bar{D}\Lambda_c\pi$ and $\bar{D}^*\Lambda_c\pi$. 
In this work, we employ the effective Lagrangian approach to calculate the partial widths of the three-body and two-body decays of the hidden-charm pentaquark hadronic molecules.

  The Lagrangians describing the hadronic molecule couplings to their constituents are written as  
\begin{eqnarray}
\label{effective}
\nonumber 
\mathcal{L}_{P_{c1} \bar{D} \Sigma_{c}}&=& g_{ P_{c1} \bar{D} \Sigma_{c}} \bar{\Sigma}_{c}    \bar{D} P_{c1},  
\\ \nonumber
\mathcal{L}_{P_{c2}\bar{D} \Sigma_{c}^{\ast}}&=& g_{P_{c2}\bar{D} \Sigma_{c}^{\ast}} \bar{\Sigma}_{c\mu}^{\ast}  \bar{D} P_{c2}^{\mu},   
\\ \nonumber
\mathcal{L}_{P_{c3}\bar{D}^{\ast} \Sigma_{c}}&=&g_{P_{c3}\bar{D}^{\ast} \Sigma_{c}} \bar{\Sigma}_{c} \gamma^{5}{\gamma}^{\mu} \bar{D}^{\ast}_{\mu}P_{c3},  \\ 
 \mathcal{L}_{P_{c4}\bar{D}^{\ast} \Sigma_{c}}&=&g_{P_{c4}\bar{D}^{\ast} \Sigma_{c}}  \bar{\Sigma}_{c}  \bar{D}^{\ast}_{\mu} P_{c4}^{\mu} ,
\\ \nonumber
\mathcal{L}_{P_{c5}\bar{D}^{\ast} \Sigma_{c}^{\ast}}&=&g_{P_{c5}\bar{D}^{\ast} \Sigma_{c}^{\ast}} \bar{\Sigma}_{c}^{\ast\mu}   \bar{D}^{\ast}_{\mu}P_{c5},  
\\ \nonumber
\mathcal{L}_{P_{c6}\bar{D}^{\ast} \Sigma_{c}^{\ast}}&=&g_{P_{c6}\bar{D}^{\ast} \Sigma_{c}^{\ast}} \bar{\Sigma}_{c\mu}^{\ast}  \bar{D}^{\ast}_{\nu}\gamma^{5}{\gamma}^{\nu} P_{c6}^{\mu}, 
\\ \nonumber   
\mathcal{L}_{P_{c7}\bar{D}^{\ast} \Sigma_{c}^{\ast}}&=&g_{P_{c7}\bar{D}^{\ast} \Sigma_{c}^{\ast}} \bar{\Sigma}_{c\mu}^{\ast} \bar{D}^{\ast}_{\nu} P_{c7}^{\mu\nu} , 
\end{eqnarray}
where  $g$ with the subscript represents the molecule couplings to their constituents. { The values of these couplings are determined by the residue of the pole positions dynamically generated by the unitary amplitude via the  Lippenman-Schwinger equation, e.g.,    $T=V+VGT$, where we take the Gaussian form factor to deal with the divergence of the loop function $G$, and the contact-range  EFT approach to construct the  $\bar{D}^{(*)}\Sigma_c^{(*)}$, $\bar{D}^{(*)}\Lambda_c$, $J/\psi p$, and $\eta_c p$ coupled-channel potentials,  characterized by six unknown parameters dictated by HQSS.}   By fitting the masses and widths of   $P_{\psi}^{N}(4312)$, $P_{\psi}^{N}(4440)$, and $P_{\psi}^{N}(4457)$, we determined these six unknown parameters in Scenarios A and B. Then we predicted the masses and widths of $P_{c1}$ $\dots$ $P_{c7}$\footnote{ In Ref.~~\cite{Pan:2023hrk}, we did not explicitly present the pole position of $P_{c7}$ and relevant couplings. With the obtained parameters in Table III of Ref.~\cite{Pan:2023hrk}, we calculate the pole position of $P_{c7}$, i.e.,  $4523.62$ MeV in Scenario A and $4502.11$ MeV in Scenario B, and the corresponding couplings $1.401$ and $ 2.562$.  } as well as the corresponding  couplings $g$ in Table~V of Ref.~\cite{Pan:2023hrk}.

 The Lagrangians describing the secondary decays $\Sigma_{c}^{(\ast)}\to \Lambda_{c}\pi$ read
\begin{eqnarray}
\mathcal{L}_{\pi\Lambda_{c}{\Sigma}_{c}}&=&-i\frac{g_{\pi\Lambda_{c}{\Sigma}_{c}}}{f_{\pi}}~\bar{\Lambda}_{c}\gamma^\mu\gamma_5 \partial_\mu\pi{\Sigma}_{c},\\ \nonumber
\mathcal{L}_{\pi\Lambda_{c}{\Sigma}_{c}^*}&=&\frac{g_{\pi\Lambda_{c}{\Sigma}_{c}^*}}{f_{\pi}}~\bar{\Lambda}_{c}\partial^\mu\pi{\Sigma}_{c\mu}^*,
\end{eqnarray}
where the pion decay constant $f_{\pi}=132$ MeV.   With the decay widths of $\Sigma_{c}^{++,+,0}\to \Lambda_{c}^{+}\pi^{+,0,-}$ and $\Sigma_{c}^{\ast ++,+,0}\to \Lambda_{c}^{+}\pi^{+,0,-}$ being 1.89, 2.3, 1.83 MeV and 14.78, 17.2, 15.3 MeV, respectively~\cite{ParticleDataGroup:2022pth}, the corresponding couplings are determined as  $g_{\pi\Lambda_{c}{\Sigma}_{c}}=0.538,0.554,0.532$ and $g_{\pi\Lambda_{c}{\Sigma}_{c}^*}=0.984,1.043,1.001$, which show minor isospin breaking.

The Lagrangian describing  the $D^{\ast}$ decay into $D\pi$ is
\begin{eqnarray}
\mathcal{L}_{D D^{\ast} \pi}&=& -i g_{D D^{\ast} \pi} (D \partial^{\mu}\pi D^{\ast\dag}_{\mu}-D_{\mu}^* \partial^{\mu} \pi  D^{\dag}), 
\end{eqnarray}
where the coupling is determined to be $g_{ D^{\ast+}D^{0}\pi^{+} }=16.818$  by  reproducing the decay width of $D^{\ast+}\to D^{0} \pi^+$~\cite{ParticleDataGroup:2022pth}. According to the isospin symmetry, 
 we obtain the coupling $g_{ D^{\ast0}D^{0}\pi^{0} }=11.688$~\cite{Xie:2022hhv}.

With the above effective Lagrangians, the amplitudes of  the three-body decays $P_{c1-c4}(k_0) \to \bar{D}^{(*)}(p_3) \pi (p_2)\Lambda_c(p_1) $ in Fig.~\ref{pc14}   can be written as 
\begin{eqnarray}
\mathcal{M}_{P_{c1}\to \bar{D}\Lambda_{c}\pi}&=& i\frac{g_{P_{c1}\bar{D}\Sigma_c}g_{\Sigma_{c}\Lambda_{c}\pi}}{f_{\pi}}~\bar{u}(p_{1}) {p\!\!\!/}_{2} \gamma_{5} \frac{1}{{k\!\!\!/}_{12}-m_{\Sigma_{c}}+ i m_{\Sigma_c}\Gamma_{\Sigma_c} }u(k_{0}), \\ 
\mathcal{M}_{P_{c1}\to \bar{D}\Lambda_{c}\pi}&=& i{g_{P_{c1}\bar{D}^{\ast}\Lambda_c}g_{\bar{D}^*\bar{D}\pi}}~\bar{u}(p_{1}) p_2^{\nu} \frac{-g^{\mu\nu}+\frac{k_{23}^\mu k_{23}^{\nu}}{m_{\bar{D}^*}^2}}{k_{23}^2-m_{\bar{D}^*}^2+i m_{\bar{D}^*} \Gamma_{\bar{D}^*}}\gamma_{5}{\gamma}^{\mu}u(k_{0}), \\
\mathcal{M}_{P_{c2}\to \bar{D}\Lambda_{c}\pi}&=& i\frac{g_{P_{c2}\bar{D}\Sigma_c^{\ast}}g_{\Sigma_{c}^{\ast}\Lambda_{c}\pi}}{f_{\pi}}~\bar{u}(p_{1}) {p}_{2 \mu} \frac{{k\!\!\!/}_{12}+m_{\Sigma_{c}^{\ast}}}{k_{12}^2-m_{\Sigma_{c}^{\ast}}^2+im_{\Sigma_{c}^{\ast}}\Gamma_{\Sigma_{c}^{\ast}}}P^{\mu\nu}(k_{12})u_{\nu}(k_{0}), \\
\mathcal{M}_{P_{c2}\to \bar{D}\Lambda_{c}\pi}&=& i{g_{P_{c2}\bar{D}^{\ast}\Lambda_c}g_{\bar{D}^*\bar{D}\pi}}~\bar{u}(p_{1}) p_{2\mu} \frac{-g^{\mu\nu}+\frac{k_{23}^\mu k_{23}^{\nu}}{m_{\bar{D}^*}^2}}{k_{23}^2-m_{\bar{D}^*}^2+i m_{\bar{D}^*} \Gamma_{\bar{D}^*}}u_{\nu}(k_{0}), \\
\mathcal{M}_{P_{c3}\to \bar{D}^{\ast}\Lambda_{c}\pi}&=& i\frac{g_{P_{c3}\bar{D}^{\ast}\Sigma_c}g_{\Sigma_{c}\Lambda_{c}\pi}}{f_{\pi}}~\bar{u}(p_{1}) {p\!\!\!/}_{2} \gamma_{5} \frac{1}{{k\!\!\!/}_{12}-m_{\Sigma_{c}}+ i m_{\Sigma_c}\Gamma_{\Sigma_c}}\gamma_{5}{\gamma}^{\mu}\varepsilon_{\mu}(p_3)u(k_{0}), \\
\mathcal{M}_{P_{c3}\to \bar{D}\Lambda_{c}\pi}&=& i\frac{g_{P_{c3}\bar{D}\Sigma_c}g_{\Sigma_{c}\Lambda_{c}\pi}}{f_{\pi}}~\bar{u}(p_{1}) {p\!\!\!/}_{2} \gamma_{5} \frac{1}{{k\!\!\!/}_{12}-m_{\Sigma_{c}}+ i m_{\Sigma_c}\Gamma_{\Sigma_c}}u(k_{0}), \\
\mathcal{M}_{P_{c3}\to  \bar{D}\Lambda_c\pi}&=& i{g_{P_{c3}\bar{D}^{\ast}\Lambda_c}g_{\bar{D}^* \bar{D}\pi}}~\bar{u}(p_{1})p_2^{\nu}\frac{-g^{\mu\nu}+\frac{k_{23}^\mu k_{23}^{\nu}}{m_{\bar{D}^*}^2}}{k_{23}^2-m_{\bar{D}^*}^2+i m_{\bar{D}^*} \Gamma_{\bar{D}^*}}\gamma_{5}{\gamma}^{\mu}u(k_{0}), \\
\mathcal{M}_{P_{c4}\to \bar{D}^{\ast}\Lambda_{c}\pi}&=& i\frac{g_{P_{c4}\bar{D}^{\ast}\Sigma_c}g_{\Sigma_{c}\Lambda_{c}\pi}}{f_{\pi}}~\bar{u}(p_{1}) {p\!\!\!/}_{2} \gamma_{5} \frac{1}{{k\!\!\!/}_{12}-m_{\Sigma_{c}}+ i m_{\Sigma_c}\Gamma_{\Sigma_c}}\varepsilon_{\mu}(p_3)u^{\mu}(k_{0}),  \\
\mathcal{M}_{P_{c4}\to \bar{D}\Lambda_{c}\pi}&=& i\frac{g_{P_{c4}\bar{D}\Sigma_c^*}g_{\Sigma_{c}^*\Lambda_{c}\pi}}{f_{\pi}}~\bar{u}(p_{1})p_2^{\nu}  \frac{{k\!\!\!/}_{12}+m_{\Sigma_{c}^{\ast}}}{k_{12}^2-m_{\Sigma_{c}^{\ast}}^2+im_{\Sigma_{c}^{\ast}}\Gamma_{\Sigma_{c}^{\ast}}}P_{\mu\nu}(k_{12})u^{\mu}(k_{0}), \\
\mathcal{M}_{P_{c4}\to  \bar{D}\Lambda_c\pi}&=& i{g_{P_{c4}\bar{D}^{\ast}\Lambda_c}g_{\bar{D}^* \bar{D}\pi}}~\bar{u}(p_{1})p_2^{\nu}\frac{-g^{\mu\nu}+\frac{k_{23}^\mu k_{23}^{\nu}}{m_{\bar{D}^*}^2}}{k_{23}^2-m_{\bar{D}^*}^2+i m_{\bar{D}^*} \Gamma_{\bar{D}^*}}u^{\mu}(k_{0}), 
\end{eqnarray}
where the momenta of $k_{12}$ and $k_{23}$ are defined as $k_{12}=p_1+p_2$ and $k_{23}=p_2+p_3$, respectively, and $P^{\mu\nu}(p)=g^{\mu\nu}-\frac{1}{3}\gamma^{\mu}\gamma^{\nu}-\frac{\gamma^{\nu}p^{\mu}-\gamma^{\mu}p^{\nu}}{3m}-\frac{2p^{\mu}p^{\nu}}{3m^{2}}$.

Similarly, the amplitudes for the three-body decays $P_{c5-c7}(k_0) \to \bar{D}^{(*)}(p_3) \pi (p_2)\Lambda_c(p_1) $ in   Fig.~\ref{pc57}   can be written as 
\begin{eqnarray}
\mathcal{M}_{P_{c5}\to \bar{D}^{\ast}\Lambda_{c}\pi}&=& i\frac{g_{P_{c5}\bar{D}^{\ast}\Sigma_c^{\ast}}g_{\Sigma_{c}^{\ast}\Lambda_{c}\pi}}{f_{\pi}}~\bar{u}(p_{1}) p_{2 \mu} \frac{{k\!\!\!/}_{12}+m_{\Sigma_{c}^{\ast}}}{k_{12}^2-m_{\Sigma_{c}^{\ast}}^2+im_{\Sigma_{c}^{\ast}}\Gamma_{\Sigma_{c}^{\ast}}}P^{\mu\nu}(k_{12})\varepsilon_{\nu}(p_3)u(k_{0}),  \\
\mathcal{M}_{P_{c5}\to \bar{D}^{\ast}\Lambda_{c}\pi}&=& i\frac{g_{P_{c5}\bar{D}^{\ast}\Sigma_c}g_{\Sigma_{c}\Lambda_{c}\pi}}{f_{\pi}}~\bar{u}(p_{1}) {p\!\!\!/}_{2} \gamma_{5} \frac{1}{{k\!\!\!/}_{12}-m_{\Sigma_{c}}+ i m_{\Sigma_c}\Gamma_{\Sigma_c}}\gamma_{5}{\gamma}^{\mu}\varepsilon_{\mu}(p_3)u(k_{0}), \\
\mathcal{M}_{P_{c5}\to \bar{D}\Lambda_{c}\pi}&=& i\frac{g_{P_{c5}\bar{D}\Sigma_c}g_{\Sigma_{c}\Lambda_{c}\pi}}{f_{\pi}}~\bar{u}(p_{1}) {p\!\!\!/}_{2} \gamma_{5} \frac{1}{{k\!\!\!/}_{12}-m_{\Sigma_{c}}+ i m_{\Sigma_c}\Gamma_{\Sigma_c}}u(k_{0}), \\
\mathcal{M}_{P_{c5}\to  \bar{D}\Lambda_c\pi}&=& i{g_{P_{c5}\bar{D}^{\ast}\Lambda_c}g_{\bar{D}^* \bar{D}\pi}}~\bar{u}(p_{1})p_2^{\nu}\frac{-g^{\mu\nu}+\frac{k_{23}^\mu k_{23}^{\nu}}{m_{\bar{D}^*}^2}}{k_{23}^2-m_{\bar{D}^*}^2+i m_{\bar{D}^*} \Gamma_{\bar{D}^*}}\gamma_{5}{\gamma}^{\mu}u(k_{0}), \\
\mathcal{M}_{P_{c6}\to \bar{D}^{\ast}\Lambda_{c}\pi}&=& i\frac{g_{P_{c6}\bar{D}^{\ast}\Sigma_c^{\ast}}g_{\Sigma_{c}^{\ast}\Lambda_{c}\pi}}{f_{\pi}}~\bar{u}(p_{1}) p_{2\mu} \frac{{k\!\!\!/}_{12}+m_{\Sigma_{c}^{\ast}}}{k_{12}^2-m_{\Sigma_{c}^{\ast}}^2+im_{\Sigma_{c}^{\ast}}\Gamma_{\Sigma_{c}^{\ast}}}P^{\mu\nu}(k_{12})\varepsilon_{\rho}(p_3)\gamma_5 {\gamma}^{\rho}u_{\nu}(k_{0}),  \\
\mathcal{M}_{P_{c6}\to \bar{D}^{\ast}\Lambda_{c}\pi}&=& i\frac{g_{P_{c6}\bar{D}^{\ast}\Sigma_c}g_{\Sigma_{c}\Lambda_{c}\pi}}{f_{\pi}}~\bar{u}(p_{1}) {p\!\!\!/}_{2} \gamma_{5} \frac{1}{{k\!\!\!/}_{12}-m_{\Sigma_{c}}+ i m_{\Sigma_c}\Gamma_{\Sigma_c}}\varepsilon_{\mu}(p_3)u^{\mu}(k_{0}),  \\
\mathcal{M}_{P_{c6}\to \bar{D}\Lambda_{c}\pi}&=& i\frac{g_{P_{c6}\bar{D}\Sigma_c^*}g_{\Sigma_{c}^*\Lambda_{c}\pi}}{f_{\pi}}~\bar{u}(p_{1})p_2^{\nu}  \frac{{k\!\!\!/}_{12}+m_{\Sigma_{c}^{\ast}}}{k_{12}^2-m_{\Sigma_{c}^{\ast}}^2+im_{\Sigma_{c}^{\ast}}\Gamma_{\Sigma_{c}^{\ast}}}P_{\mu\nu}(k_{12})u^{\mu}(k_{0}), \\
\mathcal{M}_{P_{c6}\to  \bar{D}\Lambda_c\pi}&=& i{g_{P_{c6}\bar{D}^{\ast}\Lambda_c}g_{\bar{D}^* \bar{D}\pi}}~\bar{u}(p_{1})p_2^{\nu}\frac{-g^{\mu\nu}+\frac{k_{23}^\mu k_{23}^{\nu}}{m_{\bar{D}^*}^2}}{k_{23}^2-m_{\bar{D}^*}^2+i m_{\bar{D}^*} \Gamma_{\bar{D}^*}}u^{\mu}(k_{0}), \\
\mathcal{M}_{P_{c7}\to \bar{D}^{\ast}\Lambda_{c}\pi}&=& i\frac{g_{P_{c7}\bar{D}^{\ast}\Sigma_c^{\ast}}g_{\Sigma_{c}^{\ast}\Lambda_{c}\pi}}{f_{\pi}}~\bar{u}(p_{1})p_{2\mu} \frac{{k\!\!\!/}_{1}+m_{\Sigma_{c}^{\ast}}}{k_1^2-m_{\Sigma_{c}^{\ast}}^2+im_{\Sigma_{c}^{\ast}}\Gamma_{\Sigma_{c}^{\ast}}}P^{\mu\nu}(k_1)\varepsilon_{\rho}(p_3){u_{\nu}}^{\rho}(k_{0}).  
\end{eqnarray}
 
With  the above  three-body decay amplitudes,  the partial decay widths of $P_{c} \to \bar{D}^{(\ast)}\Lambda_{c}\pi$ as a function of $m_{12}^2$ and $m_{23}^2$~\cite{ParticleDataGroup:2022pth} read
\begin{equation}
d\Gamma =  \frac{1}{(2 \pi)^{3}}\frac{1}{2J+1} \frac{\overline{|\mathcal{M}|^2}}{32 m_{P_{c}}^{3}} d m_{12}^{2} d m_{23}^{2}. 
\end{equation}
with $m_{12}$ the invariant mass of $\Lambda_c \pi$, and $m_{23}$ the invariant mass of $\bar{D}^{(\ast)}\pi$ for the $P_{c} \to \bar{D}^{(\ast)}\Lambda_{c}\pi$  decays.

\section{Results and discussions}
\label{results}

\begin{table}[!h]
\caption{Masses and quantum numbers of hadrons relevant to this work~\cite{ParticleDataGroup:2022pth}. \label{mass}}
\begin{tabular}{ccc|ccc|ccc}
  \hline\hline
   Hadron & $I (J^P)$ & M (MeV) &    Hadron & $I (J^P)$ & M (MeV) &    Hadron & $I (J^P)$ & M (MeV)     \\
  \hline
        $\Sigma_{c}^{++}$ & $1(1/2^+)$ & $2453.97$ & 
      $\Sigma_{c}^{+}$ & $1(1/2^+)$ & $2452.65$ & 
      $\Sigma_{c}^{0}$ & $1(1/2^+)$ & $2453.75$ \\
  $\Sigma_{c}^{\ast ++}$ & $1(3/2^+)$ & $2518.41$ & 
      $\Sigma_{c}^{\ast+}$ & $1(3/2^+)$ & $2517.4$ & 
      $\Sigma_{c}^{\ast0}$ & $1(3/2^+)$ & $2518.48$ \\
        $\pi^{\pm}$ & $1(0^-)$ & $139.57$ & 
      $\pi^{0}$ & $1(0^-)$ & $134.98$ & 
      $\Lambda_{c}^{+}$ & $0(1/2^+)$ & $2286.46$ \\
   $\bar{D}^{0}$ & $\frac{1}{2}(0^-)$ & $1864.84$  &    $D^{-}$ & $\frac{1}{2}(0^-)$ & $1869.66$  \\
  $\bar{D}^{\ast0}$ & $\frac{1}{2}(1^-)$ & $2006.85$ &  $D^{\ast-}$ & $\frac{1}{2}(1^-)$ & $2010.26$  \\
 \hline \hline
\end{tabular}
\label{tab:masses}
\end{table}

We tabulate the masses and quantum numbers of particles relevant to this work in Table~\ref{tab:masses}. First of all,  with the couplings obtained in Table V of  Ref.~\cite{Pan:2023hrk}  we calculate the corresponding partial decay widths via the effective Lagrangians in  Eq.(\ref{effective}), and then obtain the sum of partial decay widths,   different from the imaginary parts of the pole position in Table V of  Ref.~\cite{Pan:2023hrk}. It is shown that the total decay widths of $P_{ci}$ obtained by  Eq.(\ref{effective})  account for around  $74\%$   and  $38\%$  of the imaginary parts of the pole positions for the spin-$3/2$ and spin-$1/2$ hadronic molecules in scenarios A and B.    Therefore, to avoid such differences, we rescale the couplings of the  spin-$3/2$  and spin-$1/2$ hadronic molecules  by  $86\%$ and  $62\%$ in Table V of  Ref.~\cite{Pan:2023hrk}. Since the decay channels affect little the real parts of the pole positions of the hadronic molecules, we neglect the impact of the rescaled couplings on the real parts of pole positions.      

 {  In our calculation, the main uncertainties come from the uncertainties of the couplings of the vertices in  Fig.~1 and Fig.~2. The hidden-charm pentaquark molecule couplings to their constituents are derived by the residue of the pole positions dynamically generated by the unitary amplitude, where HQSS has a $15\%$ uncertainty~\cite{Isgur:1989vq,Isgur:1990yhj}, resulting in around $4\%$ uncertainty for the couplings.   Moreover, the cutoff variation also leads to around $10\%$ uncertainty~\cite{Wu:2023rrp}. As a result, the pentaquark molecule couplings to their constituents result in around $12\%$ uncertainty.      The uncertainties of the couplings $g_{\bar{D}^*\bar{D}\pi}$ and  $g_{\Sigma_c^{(*)}\Lambda_c\pi}$ are from the experimental uncertainties~\cite{ParticleDataGroup:2022pth}, leading to around $1\%$ and $5\%$, respectively. Therefore, we estimate the 
uncertainties for the partial decay widths originating from the uncertainties of these couplings via a Monte
Carlo sampling within their $1 \sigma$ intervals.
}

In Table~\ref{tab:width1} and Table~\ref{tab:width2}, we present the partial widths of the two-body and three-body decays as well as the total widths of  $P_{ci}$  in Scenarios A and B. $P_\psi^{N}(4312)$ is predominately generated by the elastic  $\bar{D}\Sigma_c \to \bar{D}\Sigma_c$ potential.   One can see that the branching fraction  $\mathcal{B}(P_\psi^{N}(4312) \to \bar{D} \Lambda_c \pi)$ is around  $12.5\%$    for Scenario A and  $25.1\%$ for Scenario B, smaller than those of  $\mathcal{B}( P_\psi^{N}(4312) \to  J/\psi p)$ and $\mathcal{B}( P_\psi^{N}(4312) \to  \eta_c p)$ but larger than that of  $\mathcal{B}( P_\psi^{N}(4312) \to  \bar{D}\Lambda_c)$, which indicates that  $ P_\psi^{N}(4312)$ can be measured  in the $\bar{D} \Lambda_c \pi$ and $\eta_c p$ mass distributions.  $P_{c2}$ is predominately  generated by  the elastic $\bar{D}\Sigma_c^* \to \bar{D}\Sigma_c^*$ potential\footnote{In Ref.~\cite{Du:2019pij}, Du et al. predicted a similar state, named as $P_\psi^{N}(4380)$, much narrower than that observed by LHCb Collaboration in 2015~\cite{LHCb:2015yax}.   }, which only decays into $\bar{D}\Lambda_c\pi$ and $J/\psi p$. Our results indicate that the branching fractions of $\mathcal{B}[P_{c2}\to \bar{D}\Lambda_c \pi ]$ and $\mathcal{B}[P_{c2} \to J/\psi p]$ are $41.6\%$ and $58.4\%$ for Scenario A and  $17.1\%$ and $82.9\%$ for Scenario B, which indicate that $P_{c2}$ is also likely to be measured in the $\bar{D}\Lambda_c \pi$ mass distribution in the future.   In addition,  $ P_\psi^{N}(4312)$ and $P_{c2}$   decaying into $\bar{D}^*\Lambda_c \pi$   are  not allowed in our framework, implying  that  $P_\psi^{N}(4312)$ and $P_{c2}$  cannot be observed in the  $\bar{D}^*\Lambda_c \pi$ mass distribution.

\begin{table}[!h]
 \setlength{\tabcolsep}{0.1pt}
\centering
\caption{Partial widths (in units of MeV) of three-body decays $P_c^+ \to {D^{(\ast)-}} \Lambda_c^+ \pi^+$ and $P_c^+ \to \bar{D}^{(\ast)0} \Lambda_c^+ \pi^0$ and two-body decays   $P_c \to \bar{D}\Lambda_c$, $P_c \to J/\psi N$ and $P_c \to \eta_c N$  in Scenario A. 
}
\label{tab:width1}
\begin{tabular}{c  c c c c  c c c  c }
  \hline \hline
          Mode   &  ${D^{-}} \Lambda_c^+ \pi^+$ &$\bar{D}^{0} \Lambda_c^+ \pi^0$ &${D^{\ast-}} \Lambda_c^+ \pi^+$ &$\bar{D}^{\ast0} \Lambda_c^+ \pi^0$  &$\bar{D} \Lambda_c$  & $J/\psi N$  & $\eta_c N$  & Total
         \\ \hline $P_{c1}$   & $0.036\pm0.006$ &~$0.812\pm0.096$  &- &-   &~$0.004\pm0.0005$    &~$2.014\pm0.279$    &~$3.917\pm 0.543$ &~$6.783\pm0.618$
         \\  $P_{c2}$    &$2.043\pm0.310$ &~$2.235\pm0.325$  &- &- &-    &$5.995\pm0.831$  &~~~~-  &$10.273\pm0.933$
         \\  $P_{c3}$      &$ 0.814\pm0.116$ &~$1.631\pm0.188$  &~$0.002\pm0.0004$ & ~$0.035\pm0.005$  &~$0.622\pm0.086$   &~$13.560\pm1.879$    &~$1.741\pm0.241$  &~$18.505\pm1.909$
         \\  $P_{c4}$      & $0.171\pm0.025$ &$0.152\pm0.014$  &$ 0.095\pm0.014$ & $0.388\pm0.058$ &~~~~- &$0.347\pm0.048$  &~~~~- &$1.153\pm0.082$
         \\  $P_{c5}$    &$ 0.315\pm0.045$ &~ $0.722\pm0.078$ &$ 2.370\pm0.290$ &$1.970\pm0.272$ &$1.154\pm0.160$     &$8.116\pm1.126$    &$7.487\pm1.036$ &~$22.134\pm1.592$
         \\  $P_{c6}$    &$1.354\pm0.196$ & $2.407\pm0.278$  & $4.275\pm0.519$ & $3.601\pm0.474$  &~~~~-   &$9.877\pm1.369$  &~~~~-   &$21.514\pm1.576$
         \\  $P_{c7}$    &~~~~  - &~~~~ -   &$ 2.745\pm0.413$ &~ $2.499\pm0.376$  &~~~~-   &~~~~-  &~~~~-  &~$5.244\pm0.559$ \\ 
  \hline \hline
\end{tabular}
\end{table}

\begin{table}[!h]
 \setlength{\tabcolsep}{0.2pt}
\centering
\caption{Partial widths (in units of MeV) of three-body decays $P_c^+ \to {D^{(\ast)-}} \Lambda_c^+ \pi^+$ and $P_c^+ \to \bar{D}^{(\ast)0} \Lambda_c^+ \pi^0$ and two-body decays   $P_c \to \bar{D}\Lambda_c$, $P_c \to J/\psi N$ and $P_c \to \eta_c N$  in Scenario B. 
}
\label{tab:width2}
\begin{tabular}{c c c c c c c c c}
  \hline \hline
          Mode   & ${D^{-}} \Lambda_c^+ \pi^+$ & $\bar{D}^{0} \Lambda_c^+ \pi^0$ &  ${D^{\ast-}} \Lambda_c^+ \pi^+$ & $\bar{D}^{\ast0} \Lambda_c^+ \pi^0$  & $\bar{D} \Lambda_c$  &$J/\psi N$  &$\eta_c N$  &Total 
         \\ \hline $P_{c1}$     &$ 0.023\pm0.004$ &~$1.988\pm0.255$  &~~~~- &~~~~-  &$0.008\pm0.001$   &$2.208\pm0.306$  &$3.784\pm0.525$  &~$8.011\pm0.659$
         \\  $P_{c2}$    &$1.443\pm0.226$ &~$1.547\pm0.230$  &~~~~- &~~~~-    &~~~~-   &$14.259\pm1.977$  &~~~~-   &~$17.207\pm2.003$
         \\  $P_{c3}$      &  $0.410\pm0.040$ &~$2.703\pm0.350$  &~ $0.686\pm0.103$ &~$2.803\pm0.420$  &$0.932\pm0.129$   &$3.128\pm0.434$ &$0.699\pm0.097$  &~$11.361\pm0.725$
         \\  $P_{c4}$      &~ $0.052\pm0.007$ &~$0.635\pm0.091$  &~ $ 0.001\pm0.0002$ &~ $0.014\pm0.002$   &~~~~-   &$0.904\pm0.125$  &~~~~- &~$1.606\pm0.155$
         \\  $P_{c5}$    &~ $0.371\pm0.048$ &~ $2.818\pm0.382$ &~ $7.731\pm1.105$ &~$7.037\pm0.949$  &$2.181\pm0.302$  &$4.365\pm 0.605$  &$2.949\pm0.409$  &~$27.452\pm1.701$
         \\  $P_{c6}$    & $0.760\pm0.115$ &~ $1.899\pm0.211$   &~$5.668\pm0.834$ &~ $5.090\pm0.774$   &~~~~-   &$3.432\pm0.476$  &~~~~-  &~$17.849\pm1.257$
         \\  $P_{c7}$    &~~~~ - &~~~~ -   &~  $1.084\pm0.126$&~ $0.959\pm0.144$   &~~~~-   &~~~~-  &~~~~-   &~$2.043\pm0.191$ \\
  \hline \hline
\end{tabular}
\end{table}

 $P_\psi^{N}(4440)$ and  $P_\psi^{N}(4457)$ are  predominantly generated by the elastic   $\bar{D}^* \Sigma_c \to \bar{D}^* \Sigma_c$ potentials. The branching fractions of $\mathcal{B}[P_\psi(4440)^{N} \to \bar{D}\Lambda_c \pi]$ and  $\mathcal{B}[P_\psi^{N}(4440) \to \bar{D}^*\Lambda_c \pi]$ are around $13.2\%$ and $0.2\%$ for Scenario A and  $42.8\%$ and $0.9\%$ for Scenario B, which  are smaller than that  of  $\mathcal{B}[P_\psi^{N}(4440) \to J/\psi p]$ in both scenarios. In terms of the magnitude of the branching fraction, $P_\psi^{N}(4440)$ is likely  to be measured in the $\bar{D}\Lambda_c\pi$ mass distribution in future experiments.   The branching fractions $\mathcal{B}[P_\psi^{N}(4457) \to \bar{D}\Lambda_c \pi]$ and  $\mathcal{B}[P_\psi^{N}(4457) \to \bar{D}^*\Lambda_c \pi]$ are around $28.0\%$ and $41.9\%$ for Scenario A and  $27.4\%$ and $30.7\%$ for Scenario B, the magnitude of which is similar to that of $\mathcal{B}[P_\psi^{N}(4457) \to J/\psi p]$ in both scenarios, implying that $P_\psi^{N}(4457)$ can   be measured in the $\bar{D}^*\Lambda_c \pi$ and  $\bar{D}\Lambda_c \pi$  mass distributions.     In addition, our results indicate that $P_\psi^{N}(4440)$ and $P_\psi^{N}(4457)$ having the spin of $1/2$ could be observed in the  $\bar{D}\Lambda_c$ and $\eta_c p$  mass distributions. Since the branching fraction $\mathcal{B}[P_\psi^{N}(4457) \to \bar{D}^*\Lambda_c \pi]$   is  larger than that of  $\mathcal{B}[P_\psi^{N}(4440) \to \bar{D}^*\Lambda_c \pi]$ by two order of magnitudes,   only   $P_\psi^{N}(4457)$ is probably to  be observed in  the $\bar{D}^*\Lambda_c \pi$ mass distribution.   

Under HQSS, there exist three other molecules in the vicinity of the $\bar{D}^*\Sigma_c^*$ mass threshold, i.e., $P_{c5}$, $P_{c6}$, and $P_{c7}$.  The total width of $P_{c5}$ is $22.134\pm1.592$ MeV for Scenario A and $27.452\pm1.701$ MeV for Scenario B, larger than those of  $P_{\psi}^{N}(4312)$, $P_{\psi}^{N}(4440)$, and $P_{\psi}^{N}(4457)$ in both scenarios.    
The  branching fractions of $\mathcal{B}(P_{c5} \to \bar{D}^*\Lambda_c \pi)$ and $\mathcal{B}(P_{c5} \to \bar{D}\Lambda_c \pi)$ are around   $19.6\%$  and   $4.7\%$ for Scenario A and   $53.8\%$ and  $11.6\%$  for Scenario B, while the branching fractions of $\mathcal{B}(P_{c5} \to \bar{D}\Lambda_c )$,   $\mathcal{B}(P_{c5} \to J/\psi p )$, and $\mathcal{B}(P_{c5} \to \eta_c p )$  are $5.2\%$,    $36.7\%$,  and $33.8\%$ for Scenario A and  $7.9\%$,  $15.9\%$,  and  $10.7\%$ for Scenario B, implying that  the  $\bar{D}^*\Lambda_c \pi$,  $J/\psi p$, and $\eta_c p$ channels are  favorable   to search for  the  $J^{P}=1/2^-$ $\bar{D}^*\Sigma_c^*$ molecules. In both scenarios, the total width of  $P_{c6}$ is smaller than that of  $P_{c5}$. The branching fractions of $\mathcal{B}(P_{c6} \to \bar{D}\Lambda_c \pi ) $, $\mathcal{B}(P_{c6} \to \bar{D}^*\Lambda_c \pi )$, and $\mathcal{B}(P_{c6} \to J/\psi p)$ are $17.5\%$, $36.6\%$, and $45.9\%$ for Scenario A and    $14.9\%$, $60.3\%$, and $24.8\%$ for Scenario B, implying that $P_{c6}$ can be  detected in the $\bar{D}\Lambda_c \pi$, $\bar{D}^*\Lambda_c \pi$,  and $J/\psi p$ mass distributions.  
The two-body decay of  $P_{c7}$ is heavily suppressed considering only the $S$-wave potentials, leading to a width smaller than those of $P_{c5}$ and $P_{c6}$.    Our results indicate that  $P_{c7}$  dominantly  decays into $\bar{D}^*\Lambda_c \pi$, which is the most promising channel to search for the $J^{P}=5/2^-$ $\bar{D}^*\Sigma_c^*$ molecule.    Very recently, the LHCb Collaboration observed a peak structure around  $4520$ MeV with a significance of $4.5~\sigma$  in the $\bar{D}\Lambda_c \pi$ mass distribution~\cite{LHCb:2024pnt}, which could correspond to the  $J^{P}=1/2^-$ $\bar{D}^*\Sigma_c^*$ molecule or the $J^{P}=3/2^-$ $\bar{D}^*\Sigma_c^*$ molecule predicted in Ref.~\cite{Liu:2019tjn}.   

$P_{\psi}^{N}(4312)$, $P_{\psi}^{N}(4440)$, and $P_{\psi}^{N}(4457)$  have only been observed in the $J/\psi p$ mass distribution of the  $\Lambda_b$ decay~\cite{Aaij:2019vzc}, while no significant evidence of them  was observed in the $\bar{D}\Lambda_c$, $\bar{D}\Lambda_c\pi$, and $\bar{D}^*\Lambda_c\pi$ mass distributions of prompt productions~\cite{LHCb:2024pnt}. From our results, $P_{\psi}^{N}(4312)$, $P_{\psi}^{N}(4440)$, and $P_{\psi}^{N}(4457)$  are  also likely to be observed in  the $\eta_c p$ and $\bar{D}\Lambda_c \pi$ mass distributions, the   $\eta_c p$,  $\bar{D}\Lambda_c$,   and $\bar{D}\Lambda_c \pi$ mass distributions, and  the  $\eta_c p$, $\bar{D}\Lambda_c$,  $\bar{D}\Lambda_c \pi$,  and $\bar{D}^*\Lambda_c \pi$ mass distributions, respectively. As for the other four members of the multiplet,     $P_{c2}$ and  $P_{c5}$ together with  $P_{c6}$  are likely to be observed in the $J/\psi p$ and $\bar{D}\Lambda_c \pi$ mass distributions and the $\bar{D}^*\Lambda_c \pi$ and $J/\psi p$ mass distributions, and $P_{c7}$ can probably be observed in the $\bar{D}^*\Lambda_c \pi$ mass distribution. In addition, our results  indicate that 
$P_{\psi}^{N}(4312)$ and $P_{c2}$ are can not  be observed in the $\bar{D}^*\Lambda_c \pi$ mass distribution.  Such conclusions derived from the decay behaviors can help verify the molecular nature of the pentaquark states.   

\section{Summary and Discussion}
\label{sum}

Three pentaquark states $P_{\psi}^{N}(4312)$, $P_{\psi}^{N}(4440)$, and {$P_{\psi}^{N}(4457)$} discovered by the LHCb Collaboration in 2019 are nicely arranged into a complete HQSS multiplet of hadronic molecules of $\bar{D}^{(*)}\Sigma_c^{(*)}$, while the discussions on their internal structure are still ongoing. In addition to the mass spectrum of the hidden-charm pentaquark states, their decay modes are crucial to reveal their nature. Motivated by the recent measurements of the pentaquark states in the open-charm channels, we employed the contact-range EFT to systematically investigate the two-body and three-body decay modes of the$\bar{D}^{(*)}\Sigma_c^{(*)}$ molecules, where the unknown parameters are determined by fitting the masses and widths of $P_{\psi}^{N}(4312)$, $P_{\psi}^{N}(4440)$, and {$P_{\psi}^{N}(4457)$} in two scenarios which assign the spins of the  $P_{\psi}^{N}(4440)$ and {$P_{\psi}^{N}(4457)$} as either (1/2,3/2) or (3/2, 1/2).            

In our framework, the two-body partial decays of $P_c \to \bar{D}^*\Lambda_c$ and  $P_c \to \bar{D}^{(*)}\Sigma_c^{(*)}$  are absorbed into the three-body partial decays $P_c \to \bar{D}^{(*)}\Lambda_c \pi$, leading to only  the following  two-body decays $P_c \to \eta_c p$, $P_c \to J/\psi p$, and $P_c \to \bar{D}\Lambda_c$.     
We obtained the total widths as well as the two-body and three-body partial decay widths of the HQSS multiplet hadronic molecules of $\bar{D}^{(*)}\Sigma_c^{(*)}$. From the two-body partial  decays,  $P_{c2}$,  $P_{c5}$, and $P_{c6}$ are likely to be observed in the $J/\psi p$ mass distribution, and $P_{\psi}^{N}(4312)$ and  $P_{c5}$ are likely to be observed in the  $\eta_c p$ mass distribution.  From the three-body partial decays,  it is possible to observe $P_{\psi}^{N}(4312)$,  $P_{c2}$, and $P_{\psi}^{N}(4440)$ in the $\bar{D}\Lambda_c \pi$ mass distribution, and $P_{\psi}^{N}(4457)$, $P_{c5}$, $P_{c6}$, and $P_{c7}$  in the $\bar{D}^*\Lambda_c \pi$ mass distribution. In addition, our results indicate that $P_{\psi}^{N}(4312)$ and  $P_{c2}$ are not likely to be observed in   the $\bar{D}^*\Lambda_c \pi$ mass distribution.    The conclusions obtained in this work can help clarify the nature of the pentaquark states once experimental measurements of their branching fractions are performed.

\section{Acknowledgments}
 
M.Z.L is grateful to  Mao-Jun Yan and Jun-Xu Lu for useful discussions.  
This work is partly supported by the National Key R\&D Program of China under Grant No. 2023YFA1606703.  M.Z.L
acknowledges support from the National Natural Science Foundation of China under Grant No.12105007.

\bibliography{biblio}
\end{document}